\documentclass[prl,showpacs,floatfix,twocolumn]{revtex4}

\usepackage{graphicx}
\usepackage{amsmath}

\begin{document}
\title{Signatures of quantum behavior in single-qubit weak measurements}
\author{Rusko Ruskov$^1$\footnote{On leave of absence from Institute
for Nuclear Research and Nuclear Energy, Sofia BG-1784, Bulgaria},
Alexander N. Korotkov$^2$, and Ari Mizel$^1$}

\affiliation{$^1$Department of Physics and Materials Research
Institute, Penn State University, University Park, Pennsylvania
16802, U.S.A.} \affiliation{$^2$Department of Electrical
Engineering, University of California, Riverside, CA 92521-0204,
U.S.A.}

\date{\today}

\begin{abstract}
With the recent surge of interest in quantum computation, it has
become very important to develop clear experimental tests for
``quantum behavior'' in a system.  This issue has been addressed
in the past in the form of the inequalities due to Bell and those
due to Leggett and Garg.  These inequalities concern the results
of ideal projective measurements, however, which are
experimentally difficult to perform in many proposed qubit
designs, especially in many solid state qubit systems.  Here, we
show that weak continuous measurements, which are often practical
to implement experimentally, can yield particularly clear
signatures of quantum coherence, both in the measured correlation
functions and in the measured power spectrum.
\end{abstract}

 \maketitle

Since the inception of quantum mechanics, physicists have tried
to formulate a concise statement of its essential difference from
classical mechanics \cite{Bohr,Einstein}. In the past, this has
been an important endeavor mainly because it has facilitated the
everyday use of quantum mechanics: by developing intuition about
the non-classical aspects of the theory, one is better able to
apply it to explain data.  More recently, a remarkable new
benefit has emerged from the effort. It turns out that some of
the most subtle non-classical features of quantum mechanics
actually have promising applications potential.  Researchers have
proposed quantum computers and other quantum information devices
that could rely on quantum mechanical entanglement effects to
qualitatively outperform their classical counterparts in some
important tasks (see Refs. \onlinecite{Bennett,NielsenChuang}).

Concerted effort is now being directed toward the fabrication and
control of quantum systems that could constitute the components
of a quantum information device. In particular, various two-state
systems are being studied to see if they can be made to serve as
``qubits'' \cite{QComp}, the quantum computation analogue of the
classical bit.  An important practical issue arises naturally in
such research.  One would like to be able to verify that a given
candidate system is capable of exhibiting rudimentary quantum
behavior before attempting to construct an elaborate apparatus
that can execute some task.  How can an experimentalist
demonstrate that a given system is being ``quantum'' rather than
``classical''?  Sometimes, researchers present oscillatory data
and claim that their system is undergoing quantum Rabi
oscillations. While generally the most likely source of the
behavior is quantum mechanical oscillations, an alternate,
classical explanation of the oscillations is also generally
possible.

In a profound and well-known paper \cite{Bell}, J.S. Bell
formulated inequalities that must be obeyed by any local hidden
variables theory; a system that violates the inequalities is
necessarily exhibiting non-classical behavior.  Unfortunately,
the practical requirements involved in a test of Bell's theorem
can be demanding.  The system being tested must have two degrees
of freedom that can be entangled, spatially separated, and then
separately measured.  These requirements are often too stringent
to permit initial assessments of the potential of a candidate
system to serve as a qubit.

In a different context, while seeking ways to test the
predictions of quantum mechanics for macroscopic variables,
Leggett and Garg have provided ``Bell inequalities in time''
\cite{LeggettGarg}.
These inequalities are designed for testing a system with just
one degree of freedom and therefore can be much easier to apply
in the laboratory than Bell's original inequalities. However,
they still assume the ability to perform  projective measurements
on the system. In many systems that are currently under
consideration as candidate components for quantum computers,
repeated projective measurements are difficult or impossible to
perform.  Especially in solid-state systems such as
superconducting Josephson junction qubits \cite{superqubits} one
often performs only ``weak'' continuous measurements that probe
the system gradually and indicate its state after accumulating
enough information.

In this paper, we show how the ``Bell inequalities in time'' can
be formulated to test a system that is probed with weak continuous
measurements \cite{Mensky,Carmichael,WisemanMilburn,Kor-99-01}
rather than projective measurements.  We show certain advantages
to using weak rather than projective measurements (which also
proved useful in quantum optics \cite{InequalityOptics}).  We
provide an analysis of the weak measurement signal of a two-state
system, pointing out constraints that hold (under appropriate
conditions) for a classical but not for a quantum mechanical
two-state system. When experimental data violate these
constraints one therefore has a distinct signature of quantum
behavior.

To formulate weak-measurement ``Bell inequalities in time''
consider a system with a physical characteristic described by the
variable $Q(t)$. Assume that the system conforms with the
following two axioms of macrorealism \cite{LeggettGarg}: (A1)
$Q(t)$ has a well-defined value at all times and (A2) it is
possible to obtain the value of $Q(t)$ with a non-invasive
measurement.  Assume further that $Q(t)$ is bounded above and
below so that, without loss of generality, we can arrange
definitions so that $|Q(t)| \le 1$.

Choose two non-vanishing time intervals $\tau_1$ and $\tau_2$.
Then for any initial time $t$, by axiom (A1) the three numbers
$Q(t)$, $Q(t+\tau_1)$, and $Q(t+\tau_1+\tau_2)$ give
characteristics of the system at times $t$, $t+\tau_1$, and
$t+\tau_1+\tau_2$. They satisfy the inequality
\begin{eqnarray}
Q(t)Q(t+\tau_1) & + &
Q(t+\tau_1)Q(t+\tau_1+\tau_2) \nonumber \\
\label{inequality} & & { } - Q(t)Q(t+\tau_1+\tau_2) \le 1.
\end{eqnarray}
This is proved simply by maximizing the left hand side subject to
the constraints $|Q(t)|,|Q(t+\tau_1)|,|Q(t+\tau_1+\tau_2)|\le 1$
\cite{inequalitynote}.

When making weak measurements of the system, instead of obtaining
simply $Q(t)$, one collects a noisy signal
\begin{equation}
\label{I} I(t) = I_0 + \frac{\Delta I}{2} Q(t) + \xi(t)
\end{equation}
where $\xi(t)$ represents white noise \cite{Gardiner} with
vanishing time average $\left<\xi(t) \right> \equiv
\lim_{T\to\infty} \frac{1}{T} \int_{-T/2} ^{T/2}  \xi(t)dt= 0$
and with $\delta$-function correlator
$\left<\xi(t)\xi(t+\tau)\right> = \frac{S_0}{2} \delta(\tau)$;
$S_0=2e I_0$ is the spectral density and $-e$ is the electron
charge. The symbol $I(t)$ is appropriate since the measured
signal could be the current through a device like a quantum point
contact \cite{Gurvitz,Buks} (although our analysis is not limited
to this case). The background signal is $I_0$ and $\Delta I$ is
the difference between the signal associated with $Q(t) = 1$ and
$Q(t)=-1$. By appropriately averaging $I(t)$ to minimize noise,
one can obtain information about $Q(t)$. In particular, for $\tau
> 0$ the time-averaged current correlation is
\begin{eqnarray}
\label{KI}
\lefteqn{K_{I}(\tau) \equiv \left<(I(t)-I_0)(I(t+\tau)-I_0)\right> } \\
\nonumber & = & \left(\frac{\Delta I}{2}\right)^2
\left<Q(t)Q(t+\tau)\right> + \frac{\Delta I}{2}
\left<\xi(t)Q(t+\tau)\right>.
\end{eqnarray}
We have used equation (\ref{I}) and the fact that
$\left<Q(t)\xi(t+\tau)\right>=0$, for a classical or a quantum
system, as long as the state of the system does not anticipate
the future random noise in the detector. Since it is possible to
make measurements without disturbing the system (axiom (A2)),
there is no reason that any correlation has to arise between the
noise that registers in the detector and the physical
characteristic $Q(t)$ of the system being measured. In
particular, axiom (A2) implies that in principle one can arrange
that
\begin{equation}
\label{KQxi} \left<\xi(t)Q(t+\tau)\right> = 0.
\end{equation}
Indeed, even experimentally plausible detector designs exist
\cite{Tesche} that use ``ideal negative-result'' measurements
\cite{LeggettGarg} to minimize classical back-action of the
detector on the system.  Moreover, we have considered a model of
weak continuous measurement in which detector noise linearly
perturbs one of the energy parameters in the qubit Hamiltonian;
despite the back-action explicitly included in this reasonable
model, the back-action correlator in (\ref{KQxi}) still vanishes
assuming ``good symmetric oscillations'', $\left<Q(t)\right>=0$.

Using Eq.\,(\ref{KQxi}) we get $K_{I}(\tau) = (\Delta I/2)^2
\left<Q(t)Q(t+\tau)\right>$. Averaging the inequality
(\ref{inequality}) over time $t$, we conclude
\begin{equation}
\label{Kinequality}
K_{I}(\tau_1)+K_{I}(\tau_2)-K_{I}(\tau_1+\tau_2)\le
\left(\frac{\Delta I}{2} \right)^2.
\end{equation}
This is a ``Bell inequality in time'' for weak measurements. We
will show momentarily that it is violated by a quantum system.
Note that, aside from being convenient for application to
realistic experiments, this form of the inequality has a
compelling advantage over the projective measurement version.  In
the projective measurement version \cite{LeggettGarg}, one takes
an ensemble average rather than a time average of the inequality
(\ref{inequality}) in order to arrive at an inequality that is
violated in the quantum case.  In addition, one introduces
ensembles: one ensemble which is measured at times $t$,
$t+\tau_1$ to provide the ensemble average of $Q(t)Q(t+\tau_1)$,
one ensemble measured at times $t+\tau_1$, $t+\tau_1+\tau_2$ to
provide the ensemble average of $Q(t+\tau_1)Q(t+\tau_1+\tau_2)$,
and one ensemble measured at times $t$, $t+\tau_1+\tau_2$ (but
definitely not at time $t+\tau_1$) to provide the ensemble
average of $Q(t)Q(t+\tau_1+\tau_2)$. Only by refraining from
measuring at time $t+\tau_1$ can one preserve the interference
effects in a quantum system that alter the value of
$Q(t)Q(t+\tau_1+\tau_2)$ and bring about a violation of the
inequality. Because ensembles play such an important role, an
additional explicit axiom of macrorealism called ``induction''
was introduced in Refs. \cite{LeggettGarg}
to stipulate that all ensembles have identical properties.

In the case of weak continuous measurements, no ensembles need to
be introduced since a quantum system subjected to sufficiently
weak measurements still can preserve quantum coherence
\cite{Kor-99-01}; the correlators appearing in
(\ref{Kinequality}) all refer to time averages of measurements
performed continuously on a {\it single} system. (Of course, we
have not attempted to uproot all unstated assumptions of
``induction'' from our analysis; inevitably there are many
painfully ``obvious'' unstated axioms of this sort underlying any
derivation. Our point is just that we have avoided the somewhat
awkward problem of preparing ensembles of systems in the same
starting state.)

We now demonstrate that a quantum mechanical two-state system
undergoing weak measurements violates (\ref{Kinequality}) under
appropriate conditions. Consider the density matrix $\rho$ of the
system with basis chosen so that the quantity $Q(t) = \rho_{11} -
\rho_{22}$. Measurements and Hamiltonian evolution both produce
changes in $\rho$. The following stochastic equations (in
Stratonovich form) can be derived using {\it informational}
(Bayesian) analysis or by treating the measurement device as a
quantum system with finite coupling to the system being probed
and then performing sufficiently frequent projective measurements
on the measurement device \cite{Kor-99-01}
\begin{eqnarray}
\dot{\rho}_{ij} = \rho_{ij} \frac{1}{S_0}\sum_k
\rho_{kk}\left[\left(I(t) - \frac{I_k+I_i}{2}\right)(I_i-I_k)
\right.
\qquad\ && \label{Bayes} \\
\left. + \left(I(t) - \frac{I_k+I_j}{2}\right)(I_j-I_k)\right] -
\gamma_{ij}\rho_{ij} -\frac{i}{\hbar}\left[{\cal H},\rho
\right]_{ij} . && \nonumber
\end{eqnarray}
Here, ${\cal H}$ is the system Hamiltonian and $I(t)$ is the
measurement result (\ref{I}). The system decoherence rate is
$\gamma_{ij}=(1/\eta -1)\, \frac{(I_i-I_j)^2}{4S_0}$, where the
ideality $\eta$ is unity for an ideal detector like a quantum
point contact \cite{Kor-99-01,Buttiker}. The value $I_k$ is the
current through the detector when the system is in state $|k
\rangle$. For our one-qubit case of interest, $k=1,2$ and
$I_{1,2} = I_0 \pm \Delta I/2$. For simplicity, we take the
Hamiltonian to have the form ${\cal H} = (\Omega/2) (\left| 1
\right> \left< 2 \right|+\left| 2 \right> \left< 1 \right|)$ and
denote $\gamma_{12} \equiv \gamma$.
Proceeding as in \cite{Kor-osc}, stochastic equations imply
\begin{eqnarray}
\left<Q(t)Q(t+\tau)\right> & = & \left<Q^2(t)\right> e^{-\Gamma
\tau/2} (\cos\tilde{\Omega} \tau +
\frac{\Gamma}{2\tilde{\Omega}}\sin \tilde{\Omega} \tau)
\nonumber \\
& - & \left<2 \, \mathrm{Im} \rho_{12}(t) Q(t)\right> e^{-\Gamma
\tau/2} \frac{\Omega}{\tilde{\Omega}} \sin \tilde{\Omega}\tau
\label{Q-correlator}  ,
\end{eqnarray}
\begin{eqnarray}
\lefteqn{\left<\xi(t)Q(t+\tau)\right> }
\nonumber \\
& = & \frac{\Delta I}{2} (1-\left<Q^2(t)\right>) e^{-\Gamma
\tau/2} (\cos\tilde{\Omega} \tau+ \frac{\Gamma}{2\tilde{\Omega}}
\sin \tilde{\Omega} \tau)
\nonumber \\
& & {} + \frac{\Delta I}{2} \left<2 \mathrm{Im} \rho_{12}(t)
Q(t)\right> e^{-\Gamma \tau/2} \frac{\Omega}{\tilde{\Omega}} \sin
\tilde{\Omega} \tau \label{backaction}
\end{eqnarray}
where $\tilde{\Omega} = \sqrt{\Omega^2 - \Gamma^2/4}$ and the
total decoherence rate is $\Gamma = \gamma + (\Delta I)^2/4 S_0$
$=(\Delta I)^2/4S_0\eta$. We conclude that the current
correlation (\ref{KI}) in the quantum case has the form
\cite{Kor-osc,quantum-sp}
\begin{equation}
\label{KIformula} K_{I}(\tau) = \left(\frac{\Delta I}{2}\right)^2
e^{-\Gamma \tau/2} (\cos\tilde{\Omega} \tau +
\frac{\Gamma}{2\tilde{\Omega}} \sin \tilde{\Omega} \tau).
\end{equation}

The second correlator, (\ref{backaction}), shows an inevitable
back action of noise from the detector into the evolution of the
system. This ``invasiveness'' is an essential difference between
a quantum system and a macrorealistic system satisfying axiom
(A2) above. For an ideal ($\eta=1$) detector and in the weak
coupling regime $\Gamma\ll\Omega$, we find that
$\left<Q^2(t)\right>\rightarrow 1/2$ and $\left<2 \mathrm{Im}
\rho_{12}(t) Q(t)\right> \sim \Gamma/\Omega$, so that the two
correlators (\ref{Q-correlator}), (\ref{backaction}) give an
equal contribution to the total correlation function
(\ref{KIformula}).

Choosing $\tau_1 = \tau_2 = \tau\ll 1/\Gamma$ in the inequality
(\ref{Kinequality}) and using (\ref{KIformula}), we find in the
weak coupling limit, $\Gamma \ll \Omega$, that
\begin{eqnarray}
\lefteqn{K_{I}(\tau) + K_{I}(\tau) - K_{I}(2\tau)}
\nonumber \\
& = & \left(\frac{\Delta I}{2}\right)^2 \left(1 + 2(\cos\Omega
\tau - \cos^2\Omega \tau)\right) .
\end{eqnarray}
This violates the inequality (\ref{Kinequality}) provided that
$0<\cos(\Omega \tau)<1$. We get a maximum violation of
(\ref{Kinequality}) by choosing $\tau = \pi/3\Omega$; in this
case the left hand side becomes $(3/2)(\Delta I/2)^2$. When
experimental data violate inequality (\ref{Kinequality}), it
demonstrates that the sample is not a macrorealistic system being
probed by non-invasive measurements. If an experimentalist
struggles to make non-invasive measurements but finds that the
data inevitably violate (\ref{Kinequality}), this provides
evidence that the system is behaving non-classically. Naturally,
when the decoherence rate $\Gamma$ becomes large in
(\ref{KIformula}), it is no longer possible \cite{strongCoupling}
to violate the inequality (\ref{Kinequality}).

Often, instead of directly considering the correlator (\ref{KI}),
it is experimentally convenient to analyze its power spectrum
$S_I(\omega)\equiv 2 \int_{-\infty}^{\infty} d\tau K_I(\tau )
e^{i\omega\tau}$. We now derive inequalities that constrain the
area under peaks in the power spectrum. The presence of large
area peaks that violate these inequalities should be regarded as
evidence that the sample is not a macrorealistic system being
probed non-invasively. We employ a lemma that relates the
frequency filtration of the spectrum using a frequency window and
time averaging of the current using a time window:
\begin{equation}
\int_{-\infty}^{\infty} (S_I(\Omega+\omega)-S_0) f(\omega)
\frac{d\omega}{2\pi}=  \frac{1}{\pi} \langle\, \left|
J(\Omega,t)\right|^2 \, \rangle \label{lemma} ,
\end{equation}
where $J(t)\equiv (\Delta I/2) Q(t)$ is the ``pure'' signal.
Here, $f(\omega)$ is a frequency window that goes to zero as
$|\omega|$ increases. The Fourier transformed current signal is
$J(\Omega,t) = \int_{-\infty}^{\infty}\, J(t+\tau)\,
e^{i\Omega\tau} g(\tau) d\tau$
with time averaging over a time window
$g(\tau)$; it is related to the frequency window as:
$f(\omega) = \frac{1}{2\pi} \int_{-\infty}^{\infty}\, g(\tau) g^{*}(t+\tau)\,
e^{i \omega \tau}  d\tau dt$. If one chooses a Gaussian window
$f(\omega)=e^{-\omega^2/2\Delta^2}$ the lemma holds for a
Gaussian time window $g(\tau)=\sqrt{2}\Delta e^{-\tau^2\Delta^2}$.

The integral (\ref{lemma}) gives the area under a peak in the
power spectrum centered at frequency $\Omega$ provided the width
$\Delta$ of the frequency window is much larger than the peak
width $W$. Assuming that the peak is sufficiently narrow, it is
possible to have $W \ll \Delta \ll \Omega$. The right hand side
of (\ref{lemma}) involves a time average of
$\left|J(\Omega,t)\right|^2$ that is bounded above by its maximum value
$\mathrm{max}_{t} \left| J(\Omega,t)\right|^2$, attained at the
time $t_{\mathrm {max}}$. Defining the phase $\phi$ by
$J(\Omega,t_{\mathrm {max}}) = \left|J(\Omega,t_{\mathrm{max}})\right|\exp(i\phi)$
we note that
\begin{eqnarray}
\lefteqn{\left|J(\Omega,t_{\mathrm{max}})\right| =
\int_{-\infty}^{\infty}\, J(t_{\mathrm{max}}+\tau)\, e^{i(\Omega
\tau - \phi)} g(\tau) d\tau}
\nonumber \\
& \le & \int_{-\infty}^{\infty}\,
\left|J(t_{\mathrm{max}}+\tau)\right|\, \left|\cos(\Omega \tau -
\phi)\right| g(\tau) d\tau
\nonumber \\
& \le & \int_{-\infty}^{\infty}\, \frac{\Delta I}{2}
\left|\cos(\Omega \tau - \phi)\right| \sqrt{2}\Delta e^{-\Delta^2
\tau^2}d\tau  . \nonumber
\end{eqnarray}
In the above, both $t_{\mathrm {max}}$ and $\phi$ depend on the
realization of the measurement process however the final estimate
does not. In the final step, we note that $\Delta \ll \Omega$
implies $\int \left|\cos(\Omega \tau - \phi)\right| \exp(-\Delta^2
\tau^2) d\tau = (2 /\sqrt{\pi}\Delta) [1 + o(\frac{\Delta}{\Omega})]$
since the average of the rapidly
oscillating absolute value of cosine is $2/\pi$. The correction
term $o(\frac{\Delta}{\Omega})$ rapidly decreases for small
$\Delta$; it is less than $1\%$ for $\frac{\Delta}{\Omega} < 0.4$.
One concludes that
\begin{equation}
\int_{-\infty}^{\infty} (S_I(\Omega+\omega)-S_0) f(\omega)
\frac{d\omega}{2\pi}< \frac{8}{\pi^2} \left(\frac{\Delta
I}{2}\right)^2 [1 + o(\frac{\Delta}{\Omega})] \label{peakbound}
\end{equation}
is a bound on the area of any sufficiently narrow peak in the
power spectrum of a macrorealistic system probed non-invasively.
The upper limit of $8/\pi^2\left(\Delta I/2\right)^2$ cannot be
improved without assuming further restrictions on the form of
$I(t)$. To see this, note that the limit in (\ref{peakbound}) is
actually attained by quasi-periodic rectangular oscillations:
$Q(t)=Q_R(\Omega t + \varphi(t))$ where $Q_R(\theta)= 1$ for
$2n\pi < \theta < (2n+1)\pi$ and $Q_R(\theta)= -1$ for $(2n+1)\pi
< \theta < 2(n+1)\pi$, for $n=0,1,\ldots$ and where $\varphi (t)$
is a slowly fluctuating phase
 \cite{rectangular}.
The bound (\ref{peakbound}) is violated by the quantum power
spectrum \cite{Kor-osc,quantum-sp} obtained by Fourier
transforming (\ref{KIformula})
\begin{equation}
S_I(\omega ) = S_0 + \left(\frac{\Delta I}{2}\right)^2
\frac{4\Omega^2\Gamma} {(\omega^2-\Omega^2)^2+\Gamma^2\omega^2}
\label{quantum-spectrum}
\end{equation}
which has an area of $\left(\Delta I/2\right)^2$ under the peak
at frequency $\Omega$ \cite{eta}. If we assume that power
spectrum displays only a single narrow peak, which is at non-zero
frequency, then it is possible to reduce the bound
(\ref{peakbound}). Consider a peak of functional form
(\ref{quantum-spectrum}) generated by measurements of a classical
system (in this case the peak width is $W$; it is not limited
from below by $(\Delta I)^2/4S_0$). Suppose the prefactor of
$(\Delta I/2)^2$  is replaced with $(\Delta I/2)^2 K_0$ where
$K_0$ is a constant factor.  The Fourier transform of this power
spectrum is a correlation function of the form (\ref{KIformula})
with prefactor $(\Delta I/2)^2$ replaced by $(\Delta I/2)^2 K_0$.
Assuming that this is the output of a classical system, $K_0$ is
then constrained by (\ref{Kinequality}). Taking $\tau_1 = \tau_2
= \tau = \pi/3\Omega$ and assuming $W \tau \ll 1$, we find that
$K_0 \le 2/3$.  Thus, in this case
\begin{equation}
\int_{-\infty}^{\infty} (S_I(\Omega+\omega)-S_0) f(\omega)
\frac{d\omega}{2\pi} \le \frac{2}{3} \left(\frac{\Delta
I}{2}\right)^2. \label{singlepeakbound}
\end{equation}
The assumption of a single narrow peak in $S_I(\omega)$ has led
to a more stringent constraint on the relative peak area of 2/3.
One can find a classical process with a single Lorenzian peak of
area 1/2 (e.g., $Q(t)=\cos{(\Omega t + \varphi(t))}$ with slowly
varying phase $\varphi(t)$.) Thus the exact upper bound in the
case of a {\it single} Lorenzian peak is between 1/2 and 2/3.

The three constraints (\ref{Kinequality}), (\ref{peakbound}), and
(\ref{singlepeakbound}) provide powerful and convenient means of
testing the non-classicalness of a system. Rather than simply
pointing to an oscillatory signal and claiming quantum coherent
oscillations, an experimentalist can use these inequalities to
demonstrate conclusively the violation of macroscopic,
non-invasive behavior. While it is always possible that the
experimentalist is inadvertently performing invasive measurements
on a classical system, this possibility becomes increasingly
unlikely as more effort is exerted to make the measurements
non-invasive. Although we have focused here upon a single qubit,
with little modification one can apply these constraints to weak
measurements on systems with two or more degrees of freedom.

The authors gratefully acknowledge the support of the Packard
foundation (R.R. and A.M.) and of NSA/ARDA/ARO (A.K.).


\begin{thebibliography}{99}
\bibitem{inequalitynote}  This inequality is
        identical to the inequality of \cite{LeggettGarg} with the simple
        redefinition $Q(t+\tau_1) \rightarrow -Q(t+\tau_1)$;
        here we note that it applies not only under the assumption
        $Q = \pm 1$ but also under the assumption $|Q|\le 1$.
\bibitem{strongCoupling}  In particular, in the strong coupling limit, $\Gamma \gg \Omega$ ($\eta$ fixed),
the quantum Zeno effect \cite{MisraSudarshan} develops and correlator (\ref{KIformula}) corresponds
to classical telegraph noise \cite{Kor-osc,Makhlin}; then $\langle Q^2(t)\rangle=1$ and the quantum
correlation (\ref{backaction}) vanishes.
\bibitem{eta} It does not depend on $\Gamma$ and quantum efficiency $\eta$.
              However, the measurement time $\tau_m$
        to accumulate the spectrum does increase for decreasing $\eta$; in particular,
        $\tau_m \gtrsim 8 S_0/(\Delta I)^2 \eta$ for small $\eta$.
\end{thebibliography}
\end{document}